
\magnification=\magstep1
\hsize=14cm
\vskip2.5cm
\centerline{\bf Pfaffian and Determinant Solutions to}
\centerline{\bf A Discretized Toda Eqaution for $B_r, C_r$ and $D_r$}
\vskip1.0cm \centerline{by}
\vskip1.0cm
\par\vskip0.3cm
\centerline{Atsuo Kuniba\footnote\dag{
E-mail: atsuo@hep1.c.u-tokyo.ac.jp}}
\centerline{Institute of Physics, University of Tokyo}
\centerline{Komaba 3-8-1, Meguro-ku, Tokyo 153 Japan}
\par\vskip0.4cm
\centerline{Shuichi Nakamura%
\footnote{\ddag}{Present address: Hitachi, Ltd. Information Systems Division,
890 Kashimada, Saiwai-ku, Kawasaki, Kanagawa, Japan}}
\centerline{Department of Electronics and Communication Engineering}
\centerline{School of Science and Engineering}
\centerline{Waseda University, Tokyo 169 Japan}
\vskip0.3cm
\centerline{and}\par\vskip0.3cm
\centerline{Ryogo Hirota\footnote\P{
E-mail: roy@hirota.info.waseda.ac.jp}}
\centerline{Department of Information and Computer Science}
\centerline{Waseda University, Tokyo 169 Japan}
\vskip5.0cm
\centerline{\bf Abstract}
\vskip0.2cm
\par
We consider a class of
2 dimensional Toda equations on discrete space-time.
It has arisen as functional relations
in commuting family of transfer matrices in
solvable lattice models associated with any classical
simple Lie algebra $X_r$.
For $X_r = B_r, C_r$ and $D_r$,
we present the solution in terms of Pfaffians and determinants.
They may be viewed as
Yangian analogues of the classical Jacobi-Trudi formula
on Schur functions.
\vfill
\eject
\beginsection 1. Introduction

Consider the following systems of difference
equations on $T^{(a)}_m(u)$
$(m \in {\bf Z}_{\ge 0}, u \in {\bf C},
a \in \{1,2,\ldots, r \})$
$$\eqalignno{
B_r:(r \ge 2)\qquad\qquad\qquad\qquad &&\cr
T^{(a)}_m(u-1)T^{(a)}_m(u+1) &=
T^{(a)}_{m+1}(u)T^{(a)}_{m-1}(u) +
T^{(a-1)}_m(u)T^{(a+1)}_m(u)&(1{\rm a})\cr
&\qquad\qquad\qquad\qquad\qquad 1 \le a \le r-2,&\cr
T^{(r-1)}_m(u-1)T^{(r-1)}_m(u+1) &=
T^{(r-1)}_{m+1}(u)T^{(r-1)}_{m-1}(u) +
T^{(r-2)}_m(u)T^{(r)}_{2m}(u),&(1{\rm b})\cr
T^{(r)}_{2m}(u-{1\over 2})T^{(r)}_{2m}(u+{1\over 2}) &=
T^{(r)}_{2m+1}(u)T^{(r)}_{2m-1}(u) \cr
&+ T^{(r-1)}_m(u-{1\over 2})
T^{(r-1)}_m(u+{1\over 2}),&(1{\rm c})\cr
T^{(r)}_{2m+1}(u-{1\over 2})T^{(r)}_{2m+1}(u+{1\over 2}) &=
T^{(r)}_{2m+2}(u)T^{(r)}_{2m}(u) +
T^{(r-1)}_m(u)T^{(r-1)}_{m+1}(u).&(1{\rm d})\cr}
$$
$$\eqalignno{
C_r:(r \ge 2)\qquad\qquad\qquad\qquad \quad&&\cr
T^{(a)}_m(u-{1\over 2})T^{(a)}_m(u+{1\over 2}) &=
T^{(a)}_{m+1}(u)T^{(a)}_{m-1}(u) +
T^{(a-1)}_m(u)T^{(a+1)}_m(u)&(2{\rm a})\cr
&\qquad\qquad\qquad\qquad\qquad 1 \le a \le r-2,&\cr
T^{(r-1)}_{2m}(u-{1\over 2})T^{(r-1)}_{2m}(u+{1\over 2}) &=
T^{(r-1)}_{2m+1}(u)T^{(r-1)}_{2m-1}(u)& \cr
&+ T^{(r-2)}_{2m}(u)
T^{(r)}_m(u-{1\over 2})T^{(r)}_m(u+{1\over 2}),&(2{\rm b})\cr
T^{(r-1)}_{2m+1}(u-{1\over 2})T^{(r-1)}_{2m+1}(u+{1\over 2}) &=
T^{(r-1)}_{2m+2}(u)T^{(r-1)}_{2m}(u) \cr
&+ T^{(r-2)}_{2m+1}(u)
T^{(r)}_m(u)T^{(r)}_{m+1}(u),&(2{\rm c})\cr
T^{(r)}_m(u-1)T^{(r)}_m(u+1) &=
T^{(r)}_{m+1}(u)T^{(r)}_{m-1}(u) +
T^{(r-1)}_{2m}(u).&(2{\rm d})\cr}
$$
$$\eqalignno{
D_r: (r \ge 4)\qquad\qquad\qquad\qquad &&\cr
T^{(a)}_m(u-1)T^{(a)}_m(u+1) &=
T^{(a)}_{m+1}(u)T^{(a)}_{m-1}(u) +
T^{(a-1)}_m(u)T^{(a+1)}_m(u)&(3{\rm a})\cr
&\qquad\qquad\qquad\qquad\qquad 1 \le a \le r-3,&\cr
T^{(r-2)}_m(u-1)T^{(r-2)}_m(u+1) &=
T^{(r-2)}_{m+1}(u)T^{(r-2)}_{m-1}(u) \cr
&+ T^{(r-3)}_m(u)T^{(r-1)}_m(u)T^{(r)}_m(u),&(3{\rm b})\cr
T^{(a)}_{m}(u-1)T^{(a)}_{m}(u+1) &=
T^{(a)}_{m+1}(u)T^{(a)}_{m-1}(u)
+ T^{(r-2)}_m(u)\quad a = r-1, r.&(3{\rm c})\cr}
$$
($T^{(0)}_m(u) = 1$.)
We shall exclusively consider the initial condition
$T^{(a)}_0(u) = 1$ for any $1 \le a \le r$.
Then one can solve the systems (1), (2) and (3) iteratively
to express $T^{(a)}_m(u)$ in terms of
$T^{(1)}_1(u+ \hbox{shift}),\ldots ,T^{(r)}_1(u+ \hbox{shift})$.
For example,
$T^{(1)}_2(u) = T^{(1)}_1(u-1)T^{(1)}_1(u+1)-T^{(2)}_1(u)$
from (1a).
The purpose of this paper is to present the formulae that
express an arbitrary $T^{(a)}_m(u)$ ($m \ge 1$)
as a determinant or a Pfaffian of matrices
with elements 0 or $\pm T^{(b)}_1(u+\hbox{shift})$
($0 \le b \le r$).
\par
In fact, such formulae
had been partially conjectured in [KNS1],
where a set of functional relations, $T$-system,
was introduced
for the commuting family of transfer matrices
$\{ T^{(a)}_m(u) \}$ for solvable lattice models
associated with any classical simple Lie algebra
$X_r$.
In this context, eqs. (1), (2) and (3) correspond to
$X_r = B_r, C_r$ and $D_r$ cases of the $T$-system, respectively.
$T^{(a)}_m(u)$ denotes a transfer matrix (or its eigenvalue)
with spectral parameter $u$ and ``fusion type"
labeled by $a$ and $m$ [KNS1].
Our result here confirms all of the determinant conjectures
raised in section 5 of [KNS1].
Moreover it extends them
to a full solution of (1), (2) and (3), which, in general,
involves Pfaffians as well.
In the representation theoretical viewpoint,
this yields a Yangian analogue of the Jacobi-Trudi formula [Ma], i.e.,
a way to construct Yangian characters out of
those for the fundamental representations [CP].
\par
Beside the significance in the lattice model context [KNS2],
the beautiful structure in these solutions
indicate a rich content of the $T$-system
also as an example of discretized soliton equations
[AL,H1,H2,H3,H4,HTI,K,DJM,S,VF,BKP,Wa,Wi].
In fact, regarding $u$ and $m$
as continuous space-time coordinates,
one can take a suitable scaling limit where the $T$-system
becomes 2-dimensional Toda
(or Toda molecule) equation for $X_r$ [T,MOP,LS]:
$$
(\partial_u^2 - \partial_m^2)\hbox{ log } \phi_a(u,m)
= \hbox{const } \prod_{b=1}^r
\phi_b(u,m)^{-A_{a b}}. \eqno(4)$$
Here $\phi_a(u,m)$ is a scaled $T^{(a)}_m(u)$
and $A_{a b} = {2(\alpha_a \vert \alpha_b)\over
(\alpha_a \vert \alpha_a)}$ is the Cartan matrix.
In this sense, our $T$-system is a
discretization of the Toda equation
allowing determinant and Pfaffian solutions at
least for $X_r = A_r, B_r, C_r$ and $D_r$.
See also the remarks in section 6 concerning
the $T$-systems for twisted affine Lie algebras [KS].
\par
The outline of the paper is as follows.
In sections 2, 3 and 4, we present solutions to the
$B_r, C_r$ and $D_r$ cases, respectively.
Pfaffians are needed for
$T^{(r)}_m(u)$ in $C_r$ and
$T^{(r-1)}_m(u)$ and $T^{(r)}_m(u)$ in $D_r$.
In section 5, we illustrate a proof for the $C_r$ case.
The other cases can be verified quite similarly.
Section 6 is devoted to summary and discussion.
\par
Before closing the introduction, a few remarks are in order.
Firstly, the original $T$-system [KNS1] had a factor
$g^{(a)}_m(u)$ in front of the second term
in the rhs of
$T^{(a)}_m(u+{1 \over t_a})
T^{(a)}_m(u-{1 \over t_a}) = \cdots$.
Throughout this paper we shall set $g^{(a)}_m(u) = 1$.
To recover the dependence on $g^{(a)}_m(u)$
is quite easy as long as the relation
$g^{(a)}_m(u+{1 \over t_a})
g^{(a)}_m(u-{1 \over t_a}) =
g^{(a)}_{m+1}(u)
g^{(a)}_{m-1}(u)$ is satisfied
(cf [KNS1]).
Secondly,
$T$-systems (1) and (2) coincide for $r=2$
under the exchange
$T^{(1)}_m(u) \leftrightarrow T^{(2)}_m(u)$,
which reflects the Lie algebra equivalence
$B_2 \simeq C_2$.
In this case (9) and (12) yield two alternative
expressions for the same quantity.
Thirdly,
for $X_r = A_r$, the $T$-system
$T^{(a)}_m(u-1)T^{(a)}_m(u+1) =
T^{(a)}_{m+1}(u)T^{(a)}_{m-1}(u) +
T^{(a-1)}_m(u)T^{(a+1)}_m(u)$
($1 \le a \le r, T^{(0)}_m(u) = T^{(r+1)}_m(u) = 1$)
is the so-called Hirota-Miwa equation.
In the transfer matrix context, it
has been proved in [KNS1]
by using the determinant formula in [BR].
Finally,
for $X_r = B_r$, a determinant solution different from (9) has
been obtained in [KOS].
The relevant matrix there is not sparse as (7) and
the matrix elements are not necessarily $T^{(a)}_1(u)$
but contain some quadratic expressions of $T^{(r)}_1(u)$
in general.
\beginsection 2. $B_r$ Case

For any $k \in {\bf C}$, put
$$x^a_k = \cases{T^{(a)}_1(u+k) &$1 \le a \le r$,\cr
                 1              &$a=0$. \cr}\eqno(5)
$$
We introduce the infinite dimensional matrices
${\cal T} = ({\cal T}_{i j} ) _{i,j \in {\bf Z}}$
and
${\cal E} = ({\cal E}_{i j} ) _{i,j \in {\bf Z}}$
as follows.
$$\eqalign{
{\cal T}_{i j} &= \cases{
x^{{j-i \over 2}+1}_{{i+j\over 2}-1} &
if $i \in 2{\bf Z}+1$
and ${i-j \over 2} \in \{1,0,\ldots,2-r\}$,\cr
-x^{{i-j \over 2}+2r-2}_{{i+j \over 2}-1} &
if $i \in 2{\bf Z}+1$ and
${i-j \over 2} \in
\{1-r,-r,\ldots,2-2r\}$,\cr
-x^r_{r+i-{5 \over 2}} & if $i \in 2{\bf Z}$ and $j=i+2r-3$,\cr
0 &otherwise,\cr}\cr
{\cal E}_{i j} &= \cases{
\pm 1 & if $i = j - 1 \pm 1$ and $i \in 2{\bf Z}$,\cr
x^r_{i-1} & if $i = j - 1$ and $i \in 2{\bf Z} + 1$,\cr
0 & otherwise.\cr}\cr}\eqno(6)$$
For example, for $B_3$, they read
$$
({\cal T}_{i j})_{i,j \ge 1}=
\pmatrix{     x^1_0&     0&     x^2_1&      0&
             -x^2_2&     0&   -x^1_3 &      0&    -1&       \cr
                  0&     0&         0&      0&    -x^3_{5/2}&
                  0&     0&         0&     0&               \cr
                  1&     0& x^1_2&      0&         x^2_3 &  0&
               -x^2_4&   0&  -x^1_5  &\cdots                \cr
                   0& 0& 0& 0& 0& 0& -x^3_{9/2}& 0& 0&      \cr
                   0& 0& 1& 0& x^1_4& 0& x^2_5& 0& -x^2_6&  \cr
                    & &   &  &  \vdots &&&&& \ddots \cr},
\eqno(7{\rm a})
$$
$$
({\cal E}_{i j})_{i,j \ge 1}=
\pmatrix{ 0& x^3_0&  0&     0&     0&     0&     0&   \cr
          0&     1&  0&    -1&     0&     0&     0&   \cr
          0&     0&  0& x^3_2&     0&     0&     0&   \cr
          0&     0&  0&     1&     0&    -1&     0&  \cdots \cr
          0&     0&  0&     0&     0& x^3_4&     0&   \cr
          0&     0&  0&     0&     0&     1&     0&   \cr
           &      &   &    \vdots&  &      &      &\ddots \cr
         }.
\eqno(7{\rm b})
$$
As is evident from the above example,
for any $1 \le a \le r$ and $k$, the quantity $\pm x^a_k$
is contained in ${\cal T}\vert_{u \rightarrow u + \xi}$
once and only once as its matrix element.
Here $u \rightarrow u + \xi$ means the overall shift of
lower indices in accordance with (5).
For example, the shift $\xi = 1$ is necessary to
accomodate $x^1_1$ as the (1,1) element of
${\cal T}\vert_{u \rightarrow u + \xi}$.
In view of this, we shall employ the notation
${\cal T}_m(i,j,\pm x^a_k)$ to mean the
$m$ by $m$ sub-matrix of
${\cal T}\vert_{u \rightarrow u + \xi}$
whose $(i,j)$ element is
exactly $\pm x^a_k$.
This definition is unambiguous irrespective of various possible
choices of $\xi$.
For example in (7a),
$$\eqalign{
{\cal T}_3(1,1,x^1_0) &= \pmatrix{
x^1_0 & 0 & x^2_1 \cr
0     & 0 &   0 \cr
1     & 0 &  x^1_2 \cr},
\quad
{\cal T}_3(1,1,x^1_1) = \pmatrix{
x^1_1 & 0 & x^2_2 \cr
0     & 0 &   0 \cr
1     & 0 &  x^1_3 \cr}, \cr
{\cal T}_2(1,2,-x^3_{5/2}) &= \pmatrix{
0 & -x^3_{5/2} \cr
0 & x^2_3 \cr},
\quad
{\cal T}_2(1,2,-x^3_2) = \pmatrix{
0 & -x^3_2 \cr
0 & x^2_{5/2} \cr}.
}\eqno(8)
$$
We shall also use the similar notation
${\cal E}_m(i,j,\pm x^r_k)$.
With these notations our result in this section
is stated as
\proclaim Theorem 2.1.
For $m \in {\bf Z}_{\ge 1}$,
$$\eqalignno{
T^{(a)}_m(u) &= det\bigl(
{\cal T}_{2m-1}(1,1,x^a_{-m+1}) +
{\cal E}_{2m-1}(1,2,x^r_{-m+r-a+{1\over 2}})\bigr), \,
1 \le a < r, &(9{\rm a})\cr
T^{(r)}_m(u) &= (-1)^{m(m-1)/2} det\bigl(
{\cal T}_m(1,2,-x^{r-1}_{-{m\over 2}+1}) +
{\cal E}_m(1,1,x^r_{-{m\over 2}+{1\over 2}})\bigr),
&(9{\rm b})\cr}
$$
solves the $B_r$ $T$-system (1).
\par
Up to some conventional change, (9a) in the above
had been conjectured in
eq.(5.6) of [KNS1].
The formula (9b) is new.
\beginsection 3. $C_r$ Case

Here we introduce the inifinite dimensional
matrix ${\cal T}$ by
$$
{\cal T}_{i j} = \cases{
x^{j-i+1}_{{i+j\over 2}-1} & if $i - j \in \{1,0, \ldots, 1-r\}$,\cr
- x^{i-j+2r+1}_{{i+j\over 2}-1} & if $i - j \in
\{-1-r,-2-r, \ldots, -1-2r\}$,\cr
0 & otherwise. \cr}
\eqno(10)$$
For example, for $C_2$, it reads
$$({\cal T}_{i j})_{i,j \ge 1} = \pmatrix{
x^1_0 &  x^2_{1/2} & 0 & -x^2_{3/2} & -x^1_2 & -1 & 0 & 0 &    \cr
1 &  x^1_1  &  x^2_{3/2} & 0 & -x^2_{5/2} & -x^1_3 & -1 & 0 & \cdots \cr
0 & 1 &  x^1_2  &  x^2_{5/2} & 0 & -x^2_{7/2} & -x^1_4 & -1 &  \cr
0 & 0 & 1 & x^1_3 & x^2_{7/2} & 0 & -x^2_{9/2} & -x^1_5 & \cr
&&& & \vdots &&&& \ddots \cr}.
\eqno(11)$$
We keep the same notations (5) and
${\cal T}_m(i,j,\pm x^a_k)$ ($1 \le a \le r$)
as in section 2.
Note that ${\cal T}_m(1,2,-x^r_k)$ is an anti-symmetric matrix
for any $m$.
Our result in this section is
\proclaim Theorem 3.1.
For $m \in {\bf Z}_{\ge 1}$,
$$\eqalignno{
T^{(a)}_m(u) &= det\bigl(
{\cal T}_m(1,1,x^a_{-{m\over 2}+{1\over 2}}) \bigr)
\quad 1 \le a < r, &(12{\rm a})\cr
T^{(r)}_m(u) &= (-1)^m pf \bigl(
{\cal T}_{2m}(1,2,-x^r_{-m+1}) \bigr), &(12{\rm b})\cr}
$$
solves the $C_r$ $T$-system (2).
\par
The expression (12a) is essentially the conjecture
(5.10) in [KNS1].
The Pfaffian formula (12b) is new.
In proving the theorem in section 5, we will also
establish the relations
$$\eqalignno{
T^{(r)}_m(u-{1\over 2})T^{(r)}_m(u+{1\over 2}) &=
det \bigl( {\cal T}_{2m}(1,1,x^r_{-m+{1\over 2}}) \bigr),
&(13{\rm a})\cr
T^{(r)}_m(u) T^{(r)}_{m+1}(u) &=
det \bigl( {\cal T}_{2m+1}(1,1,x^r_{-m}) \bigr).
&(13{\rm b})\cr}$$
\beginsection 4. $D_r$ Case

Here we define the infinite dimensional matrices
${\cal T}$ and ${\cal E}$ by
$$\eqalignno{
{\cal T}_{i j} &= \cases{
x^{{j-i \over 2}+1}_{{i+j\over 2}-1} &
if $i \in 2{\bf Z}+1$ and
${i-j\over 2} \in \{1,0,\ldots,3-r\}$,\cr
-x^{r-1}_{i+j-1\over 2}&if $i \in 2{\bf Z}+1$ and
${i-j\over 2}={5\over 2}-r$,\cr
-x^{r}_{i+j-3\over 2}&if $i \in 2{\bf Z}+1$ and
${i-j\over 2}={3\over 2}-r$,\cr
-x^{{i-j\over 2}+2r-3}_{{i+j\over 2}-1}
&if $i \in 2{\bf Z}+1$ and
${i-j\over 2} \in \{1-r,-r,\ldots,3-2r\}$,\cr
0&otherwise,\cr}
&(14{\rm a})\cr
{\cal E}_{i j} &= \cases{
\pm 1 &if $i=j-2 \pm 2$ and $i \in 2{\bf Z}$,\cr
x^{r-1}_i&if $i=j-3$ and $i \in 2{\bf Z}$,\cr
x^{r}_{i-2}&if $i=j-1$ and $i \in 2{\bf Z}$,\cr
0&otherwise.\cr}&(14{\rm b})\cr}$$
For example, for $D_4$, they read
$$
({\cal T}_{i j})_{i,j \ge 1} = \pmatrix{
x^1_0&      0&    x^2_1&
   -x^{3}_2&        0&    -x^{4}_2& -x^{2}_3&
        0& -x^{1}_4&             0&    -1&
                       \cr
    0&         0&            0&
            0&        0&           0&
         0&         0&          0&             0&     0&
                          \cdots \cr
    1&      0&    x^1_2&                0&x^{2}_3&  -x^{3}_4&
        0&  -x^{4}_4& -x^{2}_5&         0&-x^1_{6}&
                     \cr
    0&      0&            0&            0&
            0&        0&           0&
      0&         0&          0&          0&
                   \cr
&      &           &              &
      &         &  \vdots          &          &
         &         &      &
                      \ddots \cr
},\eqno(15{\rm a})
$$
$$
({\cal E}_{i j})_{i,j \ge 1} =
\pmatrix{
         0&        0&        0&          0&
 0&         0&          0&
         0&        0&   &   \cr
         0&        1& x^4_{0}&         0& x^{3}_{2}&       -1&
          0&
         0&        0&   \ldots&   \cr
         0&        0&        0&          0&            0&
       0&          0&
         0&        0&   &   \cr
         0&        0&        0&          1&    x^4_{2}&
        0& x^{3}_{4}&
       -1&          0&  &  \cr
   &        &         &     &             &     \vdots
    &           &
         &    & \ddots   \cr
}. \eqno(15{\rm b})
$$
We keep the same notations (5),
${\cal T}_m(i,j,\pm x^a_k)$ ($1 \le a \le r-2$)
and ${\cal T}_m(i,j,-x^a_k), {\cal E}_m(i,j,x^a_k)$ ($a = r-1, r$)
as in section 2.
Our result in this section is
\proclaim Theorem 4.1.
For $m \in {\bf Z}_{\ge 1}$,
$$\eqalignno{
&T^{(a)}_m(u) = det\bigl(
{\cal T}_{2m-1}(1,1,x^a_{-m+1}) +
{\cal E}_{2m-1}(2,3,x^r_{-m-r+a+4}) \bigr), \,\,
1 \le a \le r-2, &(16{\rm a})\cr
&T^{(r-1)}_m(u) = pf \bigl(
{\cal T}_{2m}(2,1,-x^{r-1}_{-m+1})+
{\cal E}_{2m}(1,2,x^{r-1}_{-m+1}) \bigr),
&(16{\rm b})\cr
&T^{(r)}_m(u) = (-1)^m pf \bigl(
{\cal T}_{2m}(1,2,-x^r_{-m+1})+
{\cal E}_{2m}(2,1,x^r_{-m+1}) \bigr),
&(16{\rm c})\cr}
$$
solves the $D_r$ $T$-system (3).
\par
The matrices in (16b,c) are indeed anti-symmetric.
Eq.(16a) is essentially the conjecture (5.15) in [KNS1].
The Pfaffian formulae (16b,c) are new.
By using them one can show the relations
$$\eqalignno{
&T^{(r-1)}_m(u)T^{(r)}_m(u) = (-1)^m
det \bigl( {\cal T}_{2m}(1,1,-x^{r-1}_{-m+1})+
{\cal E}_{2m}(2,2,x^r_{-m+1}) \bigr),
&(17{\rm a})\cr
&T^{(r-1)}_m(u+1)T^{(r)}_m(u-1) = (-1)^m
det \bigl( {\cal T}_{2m}(1,1,-x^{r}_{-m})+
{\cal E}_{2m}(2,2,x^{r-1}_{-m+2}) \bigr),
&(17{\rm b})\cr
&T^{(r-1)}_{m+1}(u)T^{(r)}_m(u-1) = (-1)^{m+1}
det \bigl( {\cal T}_{2m+1}(1,1,-x^{r-1}_{-m})+
{\cal E}_{2m+1}(2,2,x^r_{-m}) \bigr),
&(17{\rm c})\cr
&T^{(r-1)}_m(u+1)T^{(r)}_{m+1}(u) = (-1)^m
det \bigl( {\cal T}_{2m+1}(2,1,x^{r-2}_{-m+1})+
{\cal E}_{2m+1}(1,1,x^r_{-m}) \bigr).
&(17{\rm d})\cr}$$
The proof of (17) is analogous to that of (13), which
will be explained in the next section.
\beginsection 5. Proof of Theorem 3.1

Here we shall outline the proof of theorem 3.1, namely
$C_r$ $T$-system (2) starting from (12).
As it turns out, all of the three term relations
in (2) reduce to Jacobi's identity:
$$
D{1 \atopwithdelims
 \lbrack \rbrack 1}
D{n \atopwithdelims
 \lbrack \rbrack n} =
D D{1,n \atopwithdelims
 \lbrack \rbrack 1,n} +
D{1 \atopwithdelims
 \lbrack \rbrack n}
D{n \atopwithdelims
 \lbrack \rbrack 1}. \eqno(18)
$$
Here $D$ is the determinant of any $n$ by $n$ matrix
and $D{i_i, i_2, \ldots  \atopwithdelims
\lbrack \rbrack j_1, j_2, \ldots}$
denotes its minor removing the $i_k$'s rows and
$j_k$'s columns.
\par
Let us prove (13a) first.
Taking its square and substituting (12b),
we are to show
$$
det\big({\cal T}_{2m}(1,2,-x^r_{-m+{1\over 2}}) \bigr)
det\big({\cal T}_{2m}(1,2,-x^r_{-m+{3\over 2}}) \bigr)
= \biggl(det\big({\cal T}_{2m}(1,1,-x^r_{-m+{1\over 2}})
\bigr)\biggr)^2.\eqno(19)
$$
To see this we set
$$D = det\big({\cal T}_{2m+1}(1,2,-x^r_{-m+{1\over 2}}) \bigr)
= det \pmatrix{
0 & -x^r_{-m+{1\over 2}} & -x^{r-1}_{-m+1} & \cr
x^r_{-m+{1\over 2}} & 0 & -x^r_{-m+{3\over 2}} & \cdots \cr
x^{r-1}_{-m+1} & x^r_{-m+{3\over 2}} & 0 & \cr
  & \vdots & & \ddots \cr}
= 0, \eqno(20)
$$
since this is an anti-symmetric matrix with odd size.
{}From (20) it is easy to see
$$\eqalign{
D{1 \atopwithdelims
 \lbrack \rbrack 1} &=
det\big({\cal T}_{2m}(1,2,-x^r_{-m+{3\over 2}}) \bigr),\quad
D{2m+1 \atopwithdelims
 \lbrack \rbrack 2m+1} =
det\big({\cal T}_{2m}(1,2,-x^r_{-m+{1\over 2}}) \bigr),\cr
D{1 \atopwithdelims
 \lbrack \rbrack 2m+1} &=
D{2m+1 \atopwithdelims
 \lbrack \rbrack 1} =
det\big({\cal T}_{2m}(1,1,x^r_{-m+{1\over 2}}) \bigr).\cr}
\eqno(21)
$$
Thus (19) follows immediately from (21) and (18).
In taking the square root of (19), the relative sign
can be fixed by
comparing the coefficients of
$x^r_{-m+1/2} x^r_{-m+3/2} \cdots x^r_{m-1/2}$ on both sides,
which agrees with (13a).
The relation (13b) can be shown similarly by setting
$D = det\big({\cal T}_{2m+2}(1,2,-x^r_{-m}) \bigr)$.
\par
Now we proceed to the proof of the $T$-system (2).
To show (2a), it suffices to apply (18) for
$D = det\big({\cal T}_{m+1}(1,1,x^a_{-{m \over 2}}) \bigr) =
T^{(a)}_{m+1}(u)$ and to note that
$D{1 \atopwithdelims \lbrack \rbrack 1} = T^{(a)}_m(u+{1\over 2})$,
$D{m+1 \atopwithdelims \lbrack \rbrack m+1} = T^{(a)}_m(u-{1\over 2})$,
$D{1,m+1 \atopwithdelims \lbrack \rbrack 1,m+1} = T^{(a)}_{m-1}(u)$,
$D{m+1 \atopwithdelims \lbrack \rbrack 1} = T^{(a+1)}_m(u)$ and
$D{1 \atopwithdelims \lbrack \rbrack m+1} = T^{(a-1)}_m(u)$.
Similarly (2b) (resp. (2c)) can be derived
by setting
$D = det\big({\cal T}_{2m+1}(1,1,x^{r-1}_{-m}) \bigr) =
T^{(r-1)}_{2m+1}(u)$
(resp. $D = det\big({\cal T}_{2m+2}(1,1,x^{r-1}_{-m-{1\over 2}}) \bigr) =
T^{(r-1)}_{2m+2}(u)$)
and using (13a) (resp. (13b)) to
identify
$D{2m+1 \atopwithdelims
 \lbrack \rbrack 1}$ with
$T^{(r)}_m(u-{1\over 2})T^{(r)}_m(u+{1\over 2})$
(resp. $T^{(r)}_m(u)T^{(r)}_{m+1}(u)$).
Finally to show (2d), we put
$D = det\big({\cal T}_{2m+1}(1,1,x^{r}_{-m}) \bigr)$.
Then from (12) and (13) we have
$$\eqalign{
D &= T^{(r)}_m(u)T^{(r)}_{m+1}(u), \quad
D{1,2m+1 \atopwithdelims \lbrack \rbrack 1,2m+1} =
T^{(r)}_{m-1}(u)T^{(r)}_{m}(u), \cr
D{1 \atopwithdelims \lbrack \rbrack 1} &=
T^{(r)}_{m}(u)T^{(r)}_{m}(u+1), \quad
D{2m+1 \atopwithdelims \lbrack \rbrack 2m+1} =
T^{(r)}_{m}(u-1)T^{(r)}_{m}(u), \cr
D{1 \atopwithdelims \lbrack \rbrack 2m+1} &=
T^{(r-1)}_{2m}(u), \quad
D{2m+1 \atopwithdelims \lbrack \rbrack 1} =
\bigl(T^{(r)}_{m}(u) \bigr)^2. \cr}\eqno(22)
$$
Substituting (22) into (18) (for $n=2m+1$)
and cancelling out the common factor
$\bigl(T^{(r)}_{m}(u) \bigr)^2$, we obtain (2d).
This completes the proof of theorem 3.1.
\beginsection 6. Summary and discussion

In this paper we have considered the
difference equations (1), (2) and (3), which
may be viewed as 2 dimensional Toda equations
on discrete space-time as argued in (4).
They have arisen as the
$B_r$, $C_r$ and $D_r$ cases of the $T$-system,
which are functional relations
among commuting families of transfer matrices
in the associated solvable lattice models.
Under the initial condition $T^{(a)}_0(u) = 1$ ($1 \le a \le r$),
we have given the solutions (9), (12) and (16) for
$T^{(a)}_m(u)$ with $m \in {\bf Z}_{\ge 1}$.
They are expressed
in terms of Pfaffians or
determinants of the matrices
(6), (10) and (14), which contain only
$\pm T^{(a)}_1(u+\hbox{shift})$ or $\pm 1$ as their matrix elements.
This confirms the earlier conjectures [KNS1]
and extends them to the full solutions.
\par
It will be interesting to extend a similar analysis
to the $T$-system for the exceptional
algebras $E_{6,7,8}, F_4, G_2$ [KNS1] and also
the twisted quantum affine algebras
$A^{(2)}_n, D^{(2)}_n, E^{(2)}_6$ and $D^{(3)}_4$ [KS].
In fact, the solutions to the
$A^{(2)}_n, D^{(2)}_n$ and $D^{(3)}_4$ cases can be
obtained just by imposing the ``modulo $\sigma$ relations"
((3.4) in [KS]) on the corresponding non-twisted cases
$A_n, D_n$ and $D_4$ treated in this paper.
On the other hand, to deal with the exceptional cases,
it seems necessary to introduce matrices whose
elements are some higher order
expressions in $T^{(a)}_1(u)$ 's analogous to [KOS].
\par

\beginsection Acknowledgements

One of the authors (A.K.) thanks
E. Date, L.D. Faddeev, K. Fujii,
Y. Ohta, J. Suzuki and P.B. Wiegmann
for helpful discussions.
\vfill
\eject
\beginsection References

\item{[AL]}{M.J. Ablowitz and F.J. Ladik,
Stud.Appl.Math. {\bf 55} (1976) 213; {\bf 57} (1977) 1}
\item{[BKP]}{A. Bobenko, N. Kuts and U. Pinkall,
Phys. Lett. A{\bf 177} (1993) 399}
\item{[BR]}{V.V. Bazhanov and N.Yu. Reshetikhin,
J.Phys.A:Math.Gen. {\bf 23} (1990) 1477}
\item{[CP]}{V. Chari and A. Pressley, J. reine angew. Math. {\bf 417}
(1991) 87}
\item{[DJM]}{E. Date, M. Jimbo and T. Miwa,
J.Phys.Soc.Japan, {\bf 51} (1982) 4116; 4125;
{\bf 52} (1983) 388; 761; 766}
\item{[H1]}{R. Hirota, J.Phys.Soc.Japan, {\bf 43} (1977) 1424}
\item{[H2]}{R. Hirota, J.Phys.Soc.Japan, {\bf 45} (1978) 321}
\item{[H3]}{R. Hirota, J.Phys.Soc.Japan, {\bf 50} (1981) 3785}
\item{[H3]}{R. Hirota, J.Phys.Soc.Japan, {\bf 56} (1987) 4285}
\item{[HTI]}{R. Hitota, S. Tsujimoto and T. Imai,
in {\it Future directions of Nonlinear Dynamics in
Physical and Biological Systems}, ed. P.L. Christiansen,
J.C. Eilbeck and R.D. Parmentier, Nato ASI series (1992)}
\item{[K]}{I. M. Krichever, Russian Math. Surveys {\bf 33:4}
(1978) 255}
\item{[KNS1]}{A. Kuniba, T. Nakanishi and J. Suzuki,
Int.J.Mod.Phys. {\bf A9} (1994) 5215}
\item{[KNS2]}{A. Kuniba, T. Nakanishi and J. Suzuki,
Int.J.Mod.Phys. {\bf A9} (1994) 5267}
\item{[KOS]}{A. Kuniba, Y. Ohta and J. Suzuki,
``Quantum Jacobi-Trudi and Giambelli formulae
for $U_q(B^{(1)}_r)$ from analytic Bethe ansatz",
hep-th.9506167, J. Phys. A
in press}
\item{[KS]}{A. Kuniba and J. Suzuki, J. Phys.A: Math.Gen.
{\bf 28} (1995) 711}
\item{[LS]}{A.N. Leznov and M.V. Saveliev, Lett. Math. Phys.
{\bf 3} (1979) 489}
\item{[Ma]}{I. G. Macdonald,
{\it Symmetric functions and Hall polynomials},
2nd ed., Oxford University Press, 1995}
\item{[MOP]}{A.V. Mikhailov, M.A. Olshanetsky and
A.M. Perelomov, Commun. Math. Phys. {\bf 79} (1981) 473}
\item{[S]}{Yu.B. Suris, Phys. Lett. A{\bf 145} (1990) 113;
A{\bf 156} (1991) 467}
\item{[T]}{M. Toda, {\it Theory of nonlinear lattices},
Springer, (1988)}
\item{[VF]}{A.Yu. Volkov and L.D. Faddeev, Theor. Math. Phys.
{\bf 92} (1992) 837}
\item{[Wa]}{R.S. Ward, Phys.Lett.A{\bf 199} (1995) 45}
\item{[Wi]}{P.B. Wiegmann, ``Quantum integrable models
and discrete-time classical dynamics", Talk at
Satellite Meeting of Statphys-19 at Nankai Institute of Mathematics,
Tianjin, August (1995)}
\bye